\documentclass[12pt]{iopart}

\usepackage{iopams}  
\usepackage{graphicx}

\begin{document}

\title{Solution to the Gliding Tone Problem}

\author{Brian Cowan$^1$, Andrew Morris-Costigliola$^2$ and George Nichols$^3$}

\address{
$^1$Department of Physics, Royal Holloway University of
London, Egham, TW20 0EX, UK. \\ 
$^2$Department of Community Recovery, Brisbane, Queensland, 4000, Australia. \\ 
$^3$University of Waterloo Institute for Quantum Computing, Waterloo, ON, Canada. }


\begin{abstract}
The solution is given to the classical problem of an oscillator driven by a sinusoid of steadily-varying
frequency. A closed analytical expression is obtained in the case where the Q-factor of the oscillator
is high, equivalent to the rotating wave approximation of atomic physics. In this case all independent variables of the system combine into a single parameter. The results
are compared with previous work: series and other approximations, numerical calculations, graphical solutions
and analogue simulations. Attention is paid to the distortion of the resonance -- specifically frequency
shift and amplitude attenuation -- consequent upon the finite sweep rate. The frequency shift is interpreted
as a delay in the appearance of the resonance peak; to leading order this time delay is twice the oscillator's
ring-down time. Measurements on a high-Q oscillator are consistent with the mathematical solution.
    \end{abstract}

\maketitle

\section{Introduction}
The spectral response of a system is observed, typically, by applying an excitation which is swept over
the frequency range of interest. However the frequency-time uncertainly relation imposes the requirement of 
a sufficiently long time to measure, with sufficient accuracy, the response even to  a \emph{single} frequency.
Thus sweeping the excitation frequency at a finite rate will necessarily result in a distortion of the
true response spectrum. It is therefore of considerable practical interest to know: a)  how slow the
frequency
should be swept in order for the spectrum to be distorted negligibly, b) what distortion is caused by too-rapid
a frequency sweep, and c) the extent of any resultant shift/attenuation of the resonance peak.  

We may model this situation as an oscillator driven by a ``Gliding Tone'': an excitation of steadily varying
frequency and constant amplitude. Solving the response of an oscillator to such a Gliding Tone is a non-trivial
problem, with attempts dating back at least to the 1930's. The Gliding Tone problem has
attracted attention up to the present time but to date no exact solution has been found. This paper details
an analytic solution constrained by the single assumption that the Q-factor of the oscillator be high, equivalent, essentially, to the rotating wave approximation of atomic physics.

\section{Previous approaches}
The first attempt at the Gliding Tone problem was made by Lewis~\cite{Lewis1932}
in 1932. His interest
was in the behaviour of rotating machinery upon powering up or down, where
the rotation frequency might
pass through a mechanical resonance leading, potentially, to catastrophic
results. Lewis modelled this
system as a damped harmonic oscillator with a single degree of freedom, driven
through resonance by an
oscillation whose frequency increased with time. The solution was expressed,
using operational methods,
in terms of integrals in the complex plane. Graphical and series expansion
methods were then used to
approximate the solution for different system parameters. Thus he was able
to display the distortion
of the resonance caused by the finite frequency sweep rate.

Baker~\cite{Baker1939} applied similar ideas to the problem of an unbalanced
rotor. He was interested
in the oscillating stresses as the rotation frequency passed through a resonance.
The system was modelled
similarly. He found solutions using an analogue computer (a differential
analyzer). His particular interest
was determining appropriate damping to ensure that upon passing through resonance,
the vibration amplitude
should not become too large.

The first extensive treatment of the Gliding Tone problem was by Hok~\cite{Hok1948}.
He visualized this
in electrical terms: an LCR circuit driven by a sweep-frequency oscillator.
He used the circuit-analysis
methods of Laplace transforms to approximate the response in terms of Bromwich
integrals. Then with further
approximations, he expressed the response as a (Fresnel-like) integral. A
particular feature of Hok's
approach was that he was able to express his solution in ‘universal’
form: he found that he could
combine the system parameters  into a single independent variableγ.  
This universality is not demonstrable directly from the equation of motion;
we shall see that its origin
is more subtle. 

Barber and Ursell~\cite{Barber1948} were concerned with the design of instruments
for the accurate determination
of spectra. They discussed the distortion of the resonance profile caused
by finite sweep rates, but
they were particularly interested in the related question of determining
the optimal conditions to resolve
two close resonances. Effectively, they represented the response as an integral
over the system’s Green's
function. The bulk of the paper is then devoted to considering various approximations
that lead to analytical
expressions or series expansions for the response. In the spirit of the universality
observation articulated
by Hok, Barber and Ursell state ``the form of the response is very complicated,
but that the variation
of amplitude \emph{near resonance} (our italics) depends upon a single parameter
involving the constants
of the apparatus''.

McCann and Bennett~\cite{McCann1949} extended the consideration to systems
with two degrees of freedom,
presenting solutions obtained with the aid of an analogue computer.
Macchia~\cite{Macchia1963} was concerned with un-balanced rotating machinery.
He obtained numerical solutions.

Corliss~\cite{Corliss1963} considered the resolution limits upon sweeping
through a resonance at a finite
sweep rate. In that paper the resolution is discussed in terms of a time-frequency-energy
cube -- a ``three-way
uncertainly relation''.

Cronin~\cite{Cronin1965} devoted his PhD thesis to the Gliding Tone problem.
He explored a range of approximations
to the oscillator response, considering both linear and exponential drive
frequency sweeps. The thesis
also reported results of electrical simulations of the behaviour -- analogue
computer solutions. Various
approximation schemes were used, together with ``exact'' calculations using
such approximate expressions.
A consequence of this is that certain of his results, essentially series
expansions, are not fully correct
as the expansions have not been treated in a consistent systematic way.

Pippard~\cite{Pippard1989} gives an insightful non-mathematical description
of the physics of the Gliding
Tone problem and  Shoenberg~\cite{Shoenberg1962} has given an explicit calculation of the Gliding
Tone effect for zero damping in the vicinity of the resonance.

A completely different approach was taken by Galleani and Cohen~\cite{Galleani2000}.
They confronted the
frequency/time domain aspect of the problem directly using a Wigner distribution~\cite{Wigner1932}
description.
They demonstrated that the Wigner distribution for the Gliding Tone problem
may be calculated exactly,
using functions no more complicated than trigonometric and exponential. 
It should be mentioned that Galleani and Cohen's interest in the Gliding
Tone problem was but tangential.
They were concerned with obtaining Wigner functions (and other time-frequency
distributions) for a wide
range of differential equations and, \emph{en passant}, they discovered that
the Wigner distribution
for the Gliding Tone problem could be evaluated exactly. Knowledge of the
Gliding Tone Wigner distribution
is not so helpful at a calculational level, but it does provide insight into
the nature of the Gliding
Tone response: see Section~\ref{section_wigner}.

\section{Formulation of the Problem}
We represent the Gliding Tone problem by the following differential equation:
\begin{equation} \label{difeq1}
    \frac{{{{\mathrm{d}}^2}x(t)}}{{{\mathrm{d}}{t^2}}} + \frac{{{\omega _{\mathrm{z}}}}}{Q}\frac{{{\mathrm{d}}x(t)}}{{{\mathrm{d}}t}}
+ \omega _{\mathrm{z}}^2x(t) = f{e^{{\mathrm{i}}\left( {{\omega _{\mathrm{z}}}t + {{\alpha {t^2}} \mathord{\left/
 {\vphantom {{\alpha {t^2}} 2}} \right.
 \kern-\nulldelimiterspace} 2}} \right).}}
 \end{equation}
This specifies the variables in terms of which we shall conduct the discussion. Here $x(t)$ is the scaled
displacement of the oscillator (displacement per unit mass), $Q$ is the Q-factor of the oscillator and
$\omega _{\mathrm{z}}$  the zero-dissipation resonant angular frequency. The drive is $f{e^{{\mathrm{i}}\left(
{{\omega _{\mathrm{z}}}t + {{\alpha {t^2}} \mathord{\left/
 {\vphantom {{\alpha {t^2}} 2}} \right.
 \kern-\nulldelimiterspace} 2}} \right)}}$;  the instantaneous phase of the drive is $\phi(t)=\omega_\mathrm zt+\alpha t^2/2$, so that the instantaneous (angular) frequency is $\mathrm d \phi(t)/\mathrm d t=\omega_\mathrm z+\alpha t$. Thus  $\alpha$ is the rate at which the excitation 
frequency is increasing. We are considering an excitation frequency varying linearly with time, specified
so that the oscillator will be ``on-resonance'' at time $t=0$. 

For reference we also introduce the free  oscillation frequency $\omega _{\mathrm{f}}$ and the decay
time $\tau_{\mathrm{d}}$; the un-driven oscillator will ring down as 
\begin{equation} \label{freesol}
x\left( t \right) = A{e^{\left( {{\mathrm{i}}{\omega _{\mathrm{f}}} - {1 \mathord{\left/
 {\vphantom {1 {{\tau _{\mathrm{d}}}}}} \right.
 \kern-\nulldelimiterspace} {{\tau _{\mathrm{d}}}}}} \right)t}} + B{e^{\left( { - {\mathrm{i}}{\omega
_{\mathrm{f}}} - {1 \mathord{\left/
 {\vphantom {1 {{\tau _{\mathrm{d}}}}}} \right.
 \kern-\nulldelimiterspace} {{\tau _{\mathrm{d}}}}}} \right)t}},
 \end{equation}
where
\begin{equation} \label{wf}
 {\omega _{\mathrm{f}}} = {\omega _{\mathrm{z}}}\sqrt {1 - \frac{1}{{4{Q^2}}}}\hspace{15pt}\mathrm{and}\hspace{15pt}{\tau
_{\mathrm{d}}} = {{2Q} \mathord{\left/
 {\vphantom {{2Q} {{\omega _{\mathrm{z}}}.}}} \right.
 \kern-\nulldelimiterspace} {{\omega _{\mathrm{z}}}.}}
 \end{equation}
 
We note the equation of motion is a \emph{linear} differential equation; this permits the use of  the
complex representation to encode phase information.   
 
In the case where the oscillator is driven by a steady \emph{monochromatic} excitation, $fe^{{\mathrm{i}}
\omega_{\mathrm{d}} t}$, the response is 
\begin{equation*}
x\left( t \right) = \hat x\left( {{\omega _{\mathrm{d}}}} \right)f{e^{{\mathrm{i}}{\omega _{\mathrm{d}}}t}}.
\end{equation*}
Here $\hat x\left( {{\omega _{\mathrm{d}}}} \right)$ is the envelope of the response at the drive frequency:
\begin{equation*}
\hat x\left( {{\omega _{\mathrm{d}}}} \right) = \frac{1}{{\left( {\omega _{\mathrm{z}}^2 - \omega _{\mathrm{d}}^2}
\right) + {\mathrm{i}}{{{\omega _{\mathrm{z}}}{\omega _{\mathrm{d}}}} \mathord{\left/
 {\vphantom {{{\omega _{\mathrm{z}}}{\omega _{\mathrm{d}}}} Q}} \right.
 \kern-\nulldelimiterspace} Q}}}
\end{equation*}
and we have the high-Q limit of this expression

\begin{equation} \label{qs-sol}
\hat x\left( {{\omega _{\mathrm{d}}}} \right) = \frac{Q}{{{\omega _{\mathrm{z}}}}}\frac{1}{{2Q\left(
{{\omega _{\mathrm{z}}} - {\omega _{\mathrm{d}}}} \right) + {\mathrm{i}}{\omega _{\mathrm{z}}}}}.
\end{equation} 
This should be the response when the oscillator is driven by a \emph{very slowly} varying frequency.
The width of the resonance is $\Delta \omega=\omega_{\mathrm{z}}/Q=2/\tau_{\mathrm{d}} $.

If the oscillator were driven by a \emph{very rapidly} varying frequency, this would be equivalent to
a shock excitation. Here one should expect no response before the resonance is reached and a free ring-down,
Eq.~\ref{freesol}, after the resonance.

\section{Solution of the Problem}
\subsection{Green's function solution}

We use a Green's function method in order to obtain our solution to the Gliding Tone problem. Thus we
express the formal solution of Eq.~\ref{difeq1} as
\begin{equation} \label{gfsol}
x\left( t \right) = \int\limits_0^\infty  {f\left( {t - z} \right)G\left( z \right){\mathrm{d}}z}
\end{equation}
where $G(t)$ is the system Green's function and $f(t)$ is the drive. 
As stated above, our formulation of the problem involves a drive frequency varying linearly with time
 and the excitation
is written as
\begin{equation*}
f\left( t \right) = f{e^{{\mathrm{i}}\left( {{\omega _{\mathrm{z}}}t + {{\alpha {t^2}} \mathord{\left/
 {\vphantom {{\alpha {t^2}} 2}} \right.
 \kern-\nulldelimiterspace} 2}} \right)}}.
\end{equation*}

In order to find our solution we substitute this excitation into into Eq.~\ref{gfsol}, and we write
the result as
\begin{equation*}
x\left( t \right) = f{e^{{\mathrm{i}}\left( {{\omega _{\mathrm{z}}}t + {{\alpha {t^2}} \mathord{\left/
 {\vphantom {{\alpha {t^2}} 2}} \right.
 \kern-\nulldelimiterspace} 2}} \right)}}\int\limits_0^\infty  {{e^{{\mathrm{i}}\left( { - {\omega _{\mathrm{z}}}z
- \alpha tz + {{\alpha {z^2}} \mathord{\left/
 {\vphantom {{\alpha {z^2}} 2}} \right.
 \kern-\nulldelimiterspace} 2}} \right)}}G\left( z \right){\mathrm{d}}z}.
\end{equation*}
We have expressed it in this form to bring out the excitation as the pre-factor. Then by analogy with
the static case we introduce the envelope function $\hat x\left(t \right)$ so that:
\begin{equation*}
x\left( t \right) = \hat x\left( t \right)f{e^{{\mathrm{i}}\left( {{\omega _{\mathrm{z}}}t + {{\alpha
{t^2}} \mathord{\left/
 {\vphantom {{\alpha {t^2}} 2}} \right.
 \kern-\nulldelimiterspace} 2}} \right)}}
\end{equation*}where now the envelope depends on time since, for the Gliding Tone, $t$ is the independent
variable. 
Thus the envelope function, in our case, is
\begin{equation*}
\hat x\left( t \right) = \int\limits_0^\infty  {{e^{{\mathrm{i}}\left( { - {\omega _{\mathrm{z}}}z -
\alpha tz + {{\alpha {z^2}} \mathord{\left/
 {\vphantom {{\alpha {z^2}} 2}} \right.
 \kern-\nulldelimiterspace} 2}} \right)}}G\left( z \right){\mathrm{d}}z}. 
\end{equation*}

The Green's function, the response to a $\delta$-function excitation, is the free relaxation of the oscillator,
Eq.~\ref{freesol}, corresponding to the initial conditions $x\left( 0 \right) = 0, x'\left( 0 \right)
= 1$.
Thus
\begin{equation*}
G\left( t \right) =  - \theta \left( t \right)\frac{1}{{{\omega _{\mathrm{f}}}}}\sin {\omega _{\mathrm{f}}}t\;{e^{
- {t \mathord{\left/
 {\vphantom {t {{\tau _{\mathrm{d}}}}}} \right.
 \kern-\nulldelimiterspace} {{\tau _{\mathrm{d}}}}}}}.
\end{equation*}
It will be convenient, however, to express this function in complex form:
\begin{equation*} 
G\left( t \right) = \theta \left( t \right)\frac{{\mathrm{i}}}{{2{\omega _{\mathrm{f}}}}}{e^{ - {\mathrm{i}}{\omega
_{\mathrm{f}}}t}}{e^{ - {t \mathord{\left/
 {\vphantom {t {{\tau _{\mathrm{d}}}}}} \right.
 \kern-\nulldelimiterspace} {{\tau _{\mathrm{d}}}}}}} - \theta \left( t \right)\frac{{\mathrm{i}}}{{2{\omega
_{\mathrm{f}}}}}{e^{{\mathrm{i}}{\omega _{\mathrm{f}}}t}}{e^{ - {t \mathord{\left/
 {\vphantom {t {{\tau _{\mathrm{d}}}}}} \right.
 \kern-\nulldelimiterspace} {{\tau _{\mathrm{d}}}}}}}.
\end{equation*}
This brings out the important point that in the complex representation there are \emph{two} resonances;
one at $\omega = \omega_{\mathrm{f}}$ and another at $\omega = -\omega_{\mathrm{f}}$. In terms of the
complex Green's function the envelope function is then
\begin{equation}\label{x_hat}
\hat x(t)=\frac{-\mathrm i}{2\omega_\mathrm f}\int\limits_0^\infty e^{\mathrm i\left((\omega_\mathrm
f-\omega_\mathrm z)z-\alpha t z+\frac{\alpha t z^2}{2}+\mathrm i \frac{z}{\tau_\mathrm d}\right)} (1-e^{-2\mathrm
i\omega_\mathrm f z})\,\mathrm d z.
\end{equation}
Here we have the exact solution to the Gliding Tone problem (with linear glide). In order to make it
tractable, however, we shall consider the oscillator to have a high Q-factor.

\subsection{High-Q case}
When the $Q$ of the oscillator is high the distinction between $\omega _{\mathrm{f}}$ and $\omega _{\mathrm{z}}$
becomes vanishingly small (\emph{vide} Eq.~\ref{wf}). Upon neglecting this difference
\begin{equation*}
\hat x\left( t \right) = \frac{{ - {\mathrm{i}}}}{{2{\omega _{\mathrm{z}}}}}\int\limits_0^\infty  {{e^{{\mathrm{i}}\left(
{{{\alpha {z^2}} \mathord{\left/
 {\vphantom {{\alpha {z^2}} 2}} \right.
 \kern-\nulldelimiterspace} 2} - \alpha tz + {\mathrm{i}}{z \mathord{\left/
 {\vphantom {z {{\tau _{\mathrm{d}}}}}} \right.
 \kern-\nulldelimiterspace} {{\tau _{\mathrm{d}}}}}} \right)}}\left( {1 - {e^{ - 2{\mathrm{i}}{\omega
_{\mathrm{z}}}z}}} \right){\mathrm{d}}z}.
\end{equation*}
Observe that the oscillator's frequency has now vanished from the first exponential. In the second bracket
the second term is oscillating at double the ``carrier frequency''. But since we are calculating the envelope
function, whose variations must, by definition, be slow compared with the oscillator frequency, the double-frequency
oscillation  will average to zero over our observation time scale. Neglect of this is equivalent to ignoring
the negative-frequency component of the complex Green's function. Since, for a high-Q oscillator, the
resonances at $+\omega_{\mathrm{f}}$ and at $-\omega_{\mathrm{f}}$ will be well-separated and they will
not overlap, then it is perfectly permissible to discard the unwanted term. And so we obtain the envelope
function as
\begin{equation} \label{envint}
\hat x\left( t \right) =  - \frac{{\mathrm{i}}}{{2{\omega _{\mathrm{z}}}}}\int\limits_0^\infty  {{e^{{\mathrm{i}}\left(
{{{\alpha {z^2}} \mathord{\left/
 {\vphantom {{\alpha {z^2}} 2}} \right.
 \kern-\nulldelimiterspace} 2} + i{z \mathord{\left/
 {\vphantom {z {{\tau _{\mathrm{d}}}}}} \right.
 \kern-\nulldelimiterspace} {{\tau _{\mathrm{d}}}}} - \alpha tz} \right)}}{\mathrm{d}}z}.
\end{equation} 
         The oscillator frequency has completely vanished from the envelope function. 

In summary we have done two things; we have neglected the distinction between $\omega_{\mathrm{z}}$ and
$\omega_{\mathrm{f}}$, and we have discarded the resonance at negative frequencies. These are both acceptable
at high $Q$; thus we refer to this as the high-Q case.

The evaluation of Eq.~\ref{envint} is facilitated by expressing the integral in standard form thus:

\begin{equation} \label{intenv}
\hat x\left( t \right) =  - \frac{{\mathrm{i}}}{{{\omega _{\mathrm{z}}}\sqrt {2\alpha } }}{e^{{\mathrm{i}}{\gamma
^2}}}\int\limits_{{\mathrm{i}}\gamma }^\infty  {{e^{{\mathrm{i}}{y^2}}}{\mathrm{d}}y}
\end{equation}
where\begin{equation*}
\gamma  = \frac{1}{{\sqrt {2\alpha } {\tau _{\mathrm{d}}}}} + {\mathrm{i}}\sqrt {\frac{\alpha }{2}} t.
\end{equation*} 
         Observe at this stage that, apart from the normalization pre-factor, the shape of the envelope
function  depends on a single variable which we have cast as Hok's $\gamma$; the different parameters
of the system combine into this single variable. This may be written as
\begin{equation*}
\gamma  = \frac{1}{{\sqrt k}} \left( {\frac{1}{2} + {\mathrm{i}} \tau} \right)
\end{equation*}
where $k$ is Barber and Ursell's parameter given, in terms of our variables, by 
\begin{equation}\label{kdef}
k = \frac{\alpha \tau _{\mathrm{d}}^2}{2} =2\frac{\alpha Q^2}{\omega_{\mathrm{z}}^2}
\end{equation}
and $\tau$ is the dimensionless time:\begin{equation}\label{tau_def}
\tau  = \frac{{\alpha Q}}{{{\omega _{\mathrm{z}}}}}t.
\end{equation}

The integral in Eq.~\ref{intenv} is suggestive of Gauss's (complementary) error function (albeit of
complex argument).
In terms of this function the expression for the resonance envelope may be written 
\begin{equation} \label{erfcsol}
\hat x\left( t \right) = \frac{{\left( {1 - {\mathrm{i}}} \right)}}{{4{\omega _{\mathrm{z}}}}}\sqrt {\frac{\pi
}{\alpha }} \,{e^{{\mathrm{i}}{\gamma ^2}}}{\mathop{\mathrm {Erfc}}\nolimits} \left[ {{{\mathrm{i}}^{{1
\mathord{\left/
 {\vphantom {1 2}} \right.
 \kern-\nulldelimiterspace} 2}}}\gamma } \right].
\end{equation} 
         We note parenthetically that the integral may be expressed, equivalently, in terms of Fresnel's
functions $\mathop{\mathrm{Ci}} (z)$ and $ \mathop{\mathrm{Si}}(z) $~\cite{Abramowitz1964, Olver2010};
the complex Fresnel function $\mathop{\mathrm{F}} (z)$~\cite{Olver2010}; Dawson's integral ${\mathcal{F}}
(z)~\cite{Abramowitz1964, Olver2010} $; the plasma dispersion function~\cite{Fried1961} ${\mathrm{Z}}(z)$;
or the Kramp or Fadeeva function~\cite{Mikhailovsky1975,Fadeeva1954} $w(z)$.

\begin{figure}
\includegraphics[scale=1.0]{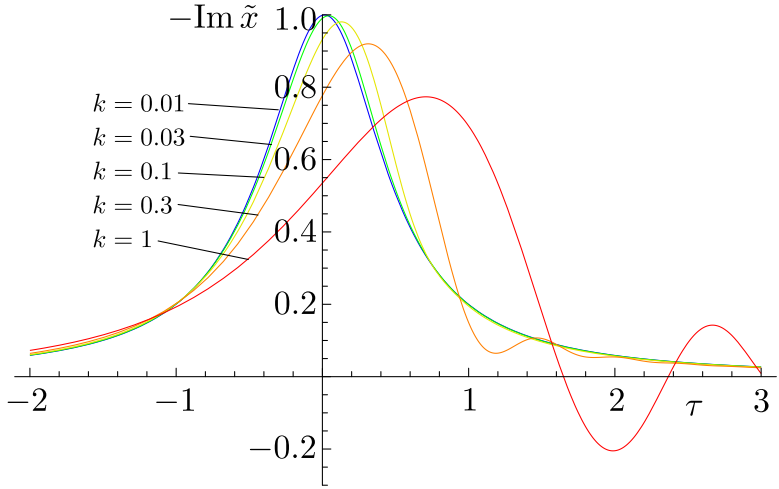}
\caption{\label{fig1}Absorption for different $k$ parameters }
\end{figure}

\begin{figure}
\includegraphics[scale=1.0]{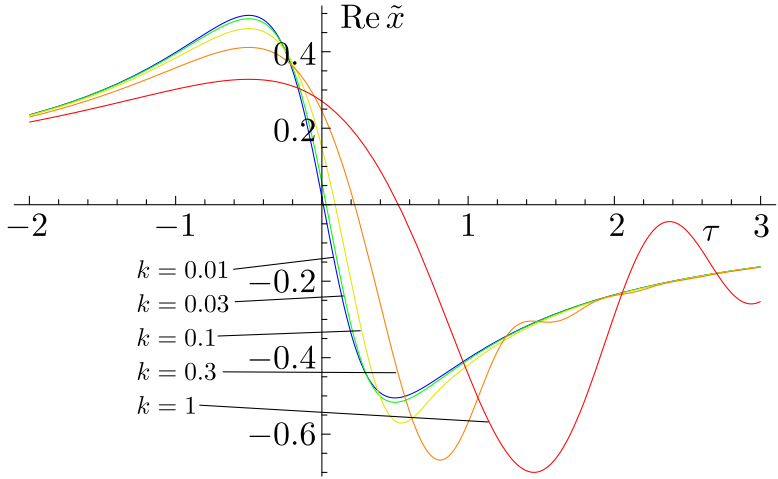}
\caption{\label{fig2}Dispersion for different $k$ parameters }
\end{figure}

\subsection{Limiting Cases}

For sufficiently slow sweep rates the envelope profile, Eq.~\ref{erfcsol}, must approach the ``static''
resonance response, Eq.~\ref{qs-sol}. We may show this to be so by expanding the error function in
Eq.~\ref{erfcsol} in inverse powers of $\gamma$. In this way we find
\begin{equation}
\hat x\left( {{\omega _{\mathrm{d}}}} \right) = \frac{Q}{{{\omega _{\mathrm{z}}}}}\frac{1}{{2Q\left(
{{\omega _{\mathrm{z}}} - {\omega _{\mathrm{d}}}} \right) + {\mathrm{i}}{\omega _{\mathrm{z}}}}}\left\{
{1 + \frac{{\mathrm{i}}}{{2{\gamma ^2}}} - \frac{3}{{4{\gamma ^4}}} - \frac{{15{\mathrm{i}}}}{{8{\gamma
^6}}} + } \right. \dots
\end{equation}
The pre-factor is the static response function and the series then shows how, with a finite sweep rate,
the response evolves from this.

For the purposes of comparison it is expedient to normalize the response so that the absorption peak
(in the quasi-static limit) has unit magnitude. Thus we define $\tilde x \left ( \tau \right)$:  
\begin{equation*}
\tilde x\left( \tau \right) = {{\hat x\left( \tau \right)} \mathord{\left /
 {\vphantom {{\hat x\left( \tau \right)} {\hat x\left( {t = 0,\;k \to 0} \right)}}} \right.
 \kern-\nulldelimiterspace} { \left \vert \hat x\left( {t = 0,\;k \to 0} \right) \right \vert}}
\end{equation*}
so that 
\begin{equation*}
\tilde x\left( \tau \right) = \frac{{\left( {1 - {\mathrm{i}}} \right)}}{4}\sqrt {\frac{{2\pi }}{k}}
\,{e^{{\mathrm{i}}{\gamma ^2}}}{\mathop{\mathrm {Erfc}}\nolimits} \left[ {\sqrt {\mathrm{i}} \,\gamma
} \right]
\end{equation*}

\begin{figure}
\includegraphics[scale=1.0]{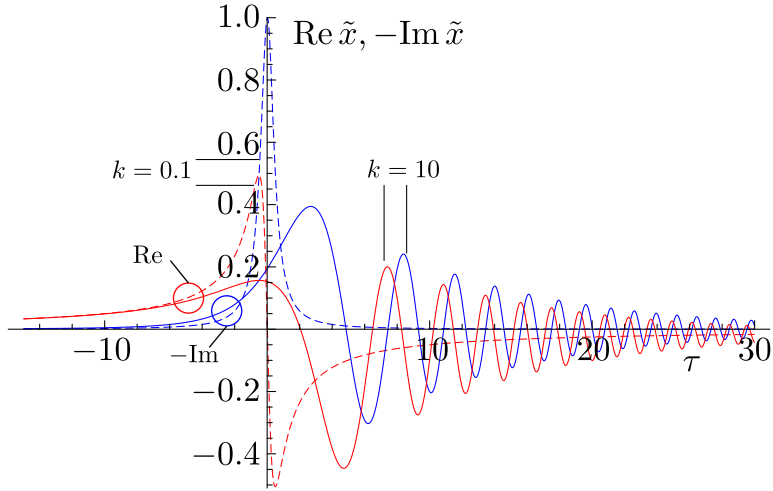}
\caption{\label{fastsweep}Absorption and dispersion for a rapid (solid lines) and a slow (dashed lines)
frequency sweep}
\end{figure}

In Figs.~1 and 2 we show the absorptive and dispersive parts of the response for different $k$ values.
For the slowest sweep rate the response is indistinguishable from that of the quasi-static case. The
first effect of a finite sweep rate is a small skewing of the resonance with a consequent delay in the
occurrence of the resonance peak. Next there is a reduction of the peak height. Then for even faster
sweep rates the resonance suffers significant distortion -- particularly \emph{after} the peak, where
the absorption can become negative.  Ultimately this evolves into no response before the resonance, and
ringing after. This is indicated in Fig.~\ref{fastsweep}. The ringing will decay with characteristic
time~$\tau_{\mathrm{d}}$. This may be seen by expanding $\tilde x \left( \tau \right)$ in powers of $\gamma$:
\begin{equation*}
\tilde x \left( \tau \right) \sim \frac{1-\mathrm{ i}}{4} \sqrt{\frac{2 \pi}{k}}e^{{\mathrm{i} / 4k}}
e^{-\tau / k}  e^{- \mathrm{i} \tau^2 /k} \left\{ 1+\mathrm{O}\left( \frac{1}{\sqrt k} \right)  \right\}
. 
\end{equation*}
The first exponential gives a simple phase factor. The second exponential gives the decay; in proper
variables this is $e^{-t/\tau_\mathrm{ d}}$. And the third exponential, in proper variables $e^{-\mathrm{i}
\alpha t^2 /2} $, gives the ringing at the ``instantaneous'' frequency deviation from $\omega_{\mathrm{z}}
$.

\subsection{Comments on previous treatments}
Our solution is a function of the reduced time $\tau$ (Eq.~\ref{tau_def}). This involves the product
of time $t$ and sweep rate $\alpha$, so reversing $t$ is equivalent to reversing $\alpha$. Thus a down
sweep will have a response that is the mirror image of an up sweep -- as expected on intuitive grounds.
However Cronin~\cite{Cronin1965} found, in his series approximations, that the mathematical forms of the
up and down sweeps were \emph{not} identical. This disagreement can be traced back to our high-Q condition:
specifically the discarding of the negative-frequency resonance. When this is not discarded it follows
that the observed resonance profile will depend upon whether the ``other'' resonance has been passed through
or not. Of course this difference will be numerically negligible in the high-Q case.
One should also note that any realistic spectral sweep would span frequencies in the vicinity of the
resonance; experimentally the other resonance would never be covered.

Hok's approach~\cite{Hok1948} involves taking an inverse Laplace transform. In a crucial step he discards
as negligible a term (involving his $\gamma_2$). This is equivalent to our discarding of the negative
frequency resonance.

\section{Resonance shift}
\subsection{Introduction}
Spectroscopy experiments often require the precise location of a resonance peak. Our results above indicate
that sweeping the response up through the resonance at a finite rate will result in the peak occurring
at a slightly higher frequency, while sweeping down will result in a slightly lower frequency.  Thus,
for example, Phillips and Gold~\cite{Phillips1969} in studying the de Haas-van Alphen effect in lead
appreciated the necessity for applying corrections for a finite sweep rate. 

\begin{figure}
\includegraphics[scale=1.0]{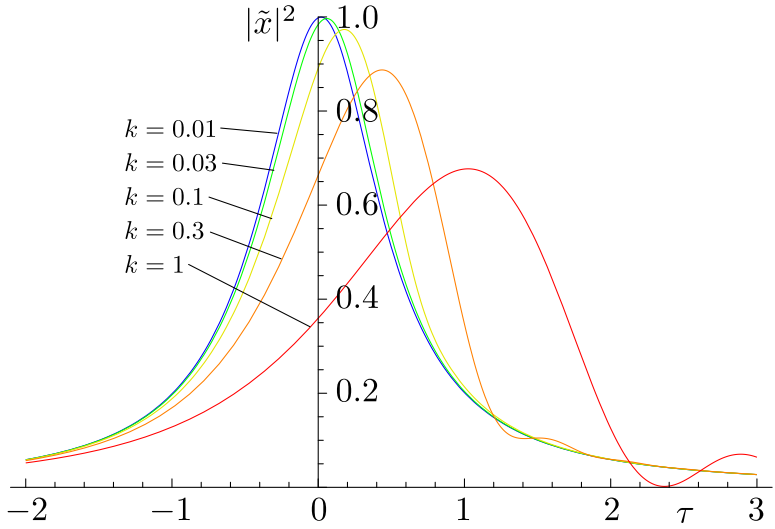}
\caption{\label{mag2-b}Variation of resonance intensity for different $k$ parameters }
\end{figure}

We shall discuss the shift in terms of the intensity of the resonance, the square magnitude response
$\left |\tilde x (\tau)\right |^2$:
\begin{equation}
|\tilde x(\tau)|^2=\frac{\pi}{4k}{  e^{-\frac{2 \tau }{k}} \mathrm{Erfc}\left[\frac{(1-\mathrm i)-2(1+ \mathrm i) \tau }{2 \sqrt{2k} }\right] \mathrm{Erfc}\left[\frac{(1+\mathrm i)-2(1- \mathrm i) \tau }{2 \sqrt{2k} }\right]}.
\end{equation}
This is shown in Fig.~\ref{mag2-b}. Series expansion of $\left |\tilde
x (\tau)\right |^2 $ shows how the intensity profile evolves from the quasi-static case.

\begin{equation}\label{x2ser}
\eqalign{
 {\left| {\tilde x\left( \tau  \right)} \right|^2} &= \frac{1}{{1 + 4{\tau ^2}}} + \frac{{16\tau }}{{{{\left(
{1 + 4{\tau ^2}} \right)}^3}}}k - \frac{{4\left( {5 - 152{\tau ^2} + 80{\tau ^4}} \right)}}{{{{\left(
{1 + 4{\tau ^2}} \right)}^5}}}{k^2} \\  &-\frac{{384\left( {7\tau  - 104{\tau ^3} + 112{\tau ^5}} \right)}}{{{{\left(
{1 + 4{\tau ^2}} \right)}^7}}}{k^3} +  \dots 
}  
\end{equation}

The first term is the quasi-static solution. The latter terms give the increasing distortion. \\

\subsection{Frequency shift}
\begin{figure}
\includegraphics[scale=0.95]{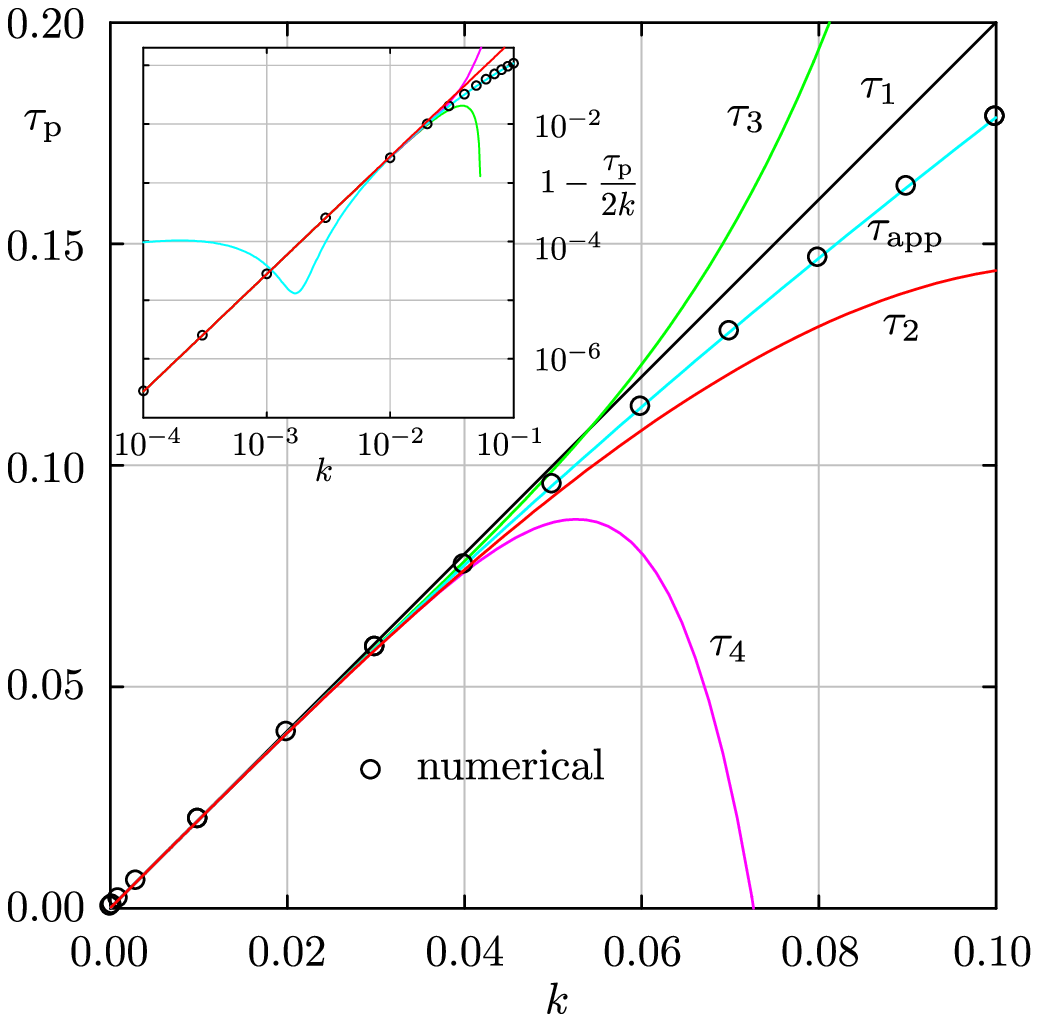}
\caption{\label{fig4}Shift of resonance frequency with increasing $k$ parameter. The black,
red, green and purple lines show the successive Taylor approximants
to the series, Eq.~\ref{sertau}: truncated after the first, third, fifth and seventh powers. }
\end{figure}

The resonance peak corresponds to the maximum of the intensity. This peak  evolves from the $\tau = 0
$ maximum as $ k$ grows from zero. In order to find the position of the peak we must solve the equation
\begin{equation} \label{dx2=0}
\frac{{{\mathrm{d}}{{\left| {\tilde x\left( \tau  \right)} \right|}^2}}}{{{\mathrm{d}}\tau }} = 0.
\end{equation}
We do this in the following neat way. We expand the derivative of Eq.~\ref{x2ser}, $d={{{\mathrm{d}}{{\left|
{\tilde x\left( \tau  \right)} \right|}^2}}}/{{{\mathrm{d}}\tau }} $
as a power series in $\tau$:
\begin{equation*}\label{serd}
\eqalign{
 d = 
&16\left( {k - 168{k^3} + 720{k^5}+\dots} \right) \\ 
&- 8\left( {1 - 252{k^2} - 5520{k^4} + 100800{k^6}+\dots} \right)\tau  \\
&- 576\left( {k - 600{k^3} + 1680{k^5}+\dots} \right){\tau ^2} \\ 
&+64\left( {1 - 1080{k^2} - 31920{k^4} + 403200{k^6}+\dots} \right){\tau ^3}
 +  \dots 
 }
\end{equation*}
and we want to find the value of $\tau $ that will make this expression zero. We may revert this series,
to give $\tau$ as a series in $d$. Then all we require is the leading term, corresponding to $d=0$. Thus
we find the (reduced) time for the occurrence of the intensity peak, $\tau_{\mathrm{p}}$, as
\begin{equation}\label{sertau}
{\tau _{\mathrm{p}}}\left(k\right) = 2k - 56{k^3} + 19104{k^5} - 15194496{k^7} +  \dots
\end{equation}
This affirms that there is no shift when $k$ vanishes and it indicates how the shift evolves with $k$.
However the successive coefficients of the series increase rapidly; this is a divergent series as indicated in Fig.~\ref{fig4}, of use only for small $k$.
The points in the figure are  found directly from numerical solution of Eq.~\ref{dx2=0}. The lines in the figure show successive
Taylor approximants to the series.

It is clear that for small $k$ the leading-term linear behaviour is dominant. However the deviation from
linearity is better-described, over a larger $k$ range, by a power series in $k^{1/2}$; this is shown by the cyan line in the figure, a least squares fit through the numerical points.
Its equation is
\begin{equation}\label{tau-app}
\tau_{\mathrm {app}}\left(k\right)  = 2k + {b_1}{k^{{3 \mathord{\left/
 {\vphantom {3 2}} \right.
 \kern-\nulldelimiterspace} 2}}} + {b_2}{k^2} + {b_3}{k^{{5 \mathord{\left/
 {\vphantom {5 2}} \right.
 \kern-\nulldelimiterspace} 2}}} + {b_4}{k^3}
\end{equation}
where
\begin{equation*}
\eqalign{
 {b_1} &=  - 0.0334937425 \\ 
 {b_2} &= \phantom{-}1.6315734132 \\ 
 {b_3} &=  - 21.535018954 \\ 
 {b_4} &= \phantom{-}31.360499596\;. \\ 
 }
\end{equation*}
This enables an accurate determination of the true resonance frequency $\omega_{\mathrm{z}}$ from the
observed resonance frequency $\omega_{\mathrm{p}}$. 

The inset to the figure, shows $1-\frac{\tau_\mathrm p}{2k}$ against $k$ with logarithmic axes. This indicates the limitation of the series approximation while demonstrating its merit of giving $\tau_\mathrm p$ with sufficient accuracy for larger
values of $k$.

The reduced time of the peak $\tau_{\mathrm{p}}$ corresponds to a real time delay $t_{\mathrm{p}}$:
\begin{equation*}\label{t_p}
t_{\mathrm{p}}=\frac{\omega_{\mathrm{z}}}{\alpha Q}\tau_{\mathrm{p}}
\end{equation*}
so there is a frequency shift $\delta \omega $ of $\alpha t_{\mathrm{p}}$ or
\begin{equation*}\label{shift}
\delta \omega=\frac{\omega_{\mathrm{z}}}{Q} \tau_{\mathrm{p}}\left(k\right).
\end{equation*}

For a slow sweep rate ($\alpha\ll\omega_\mathrm z^2/Q^2$), where we take only the leading term in the expansion: $\tau_{\mathrm{p}}=2k $,
the time of the peak's occurrence is
\begin{equation*}
t_{\mathrm{p}}=2\tau_{\mathrm{d}}.
\end{equation*}
This is telling us that (in the slow-sweep limit) \emph{the resonance peak suffers a time delay of twice
the oscillator's ring-down time}. This is intuitively reasonable, since $\tau_{\mathrm{d}} $ characterizes
the time it takes for the oscillator to settle -- the time it takes to respond to a stimulus. To this
leading order of approximation we find the frequency shift of the peak to be
\begin{equation}\label{deltaomega}
\delta \omega = 2k\Delta \omega:
\end{equation}   
the \emph{resonance shift is $2k$ times the resonance width}.

\subsection{Peak attenuation}
The normalized signal peak intensity and its evolution with increasing $k$ is found by substituting Eq.~\ref{sertau}
into Eq.~\ref{x2ser}:
\begin{equation} \label{x2pser}
\left| {\tilde x \left( {{\tau _{\mathrm{p}}}} \right)} \right|^2 = 1 - 4{k^2} + 400{k^4} - 152640{k^6}
+  \dots 
\end{equation}

\begin{figure}
\includegraphics[scale=0.95]{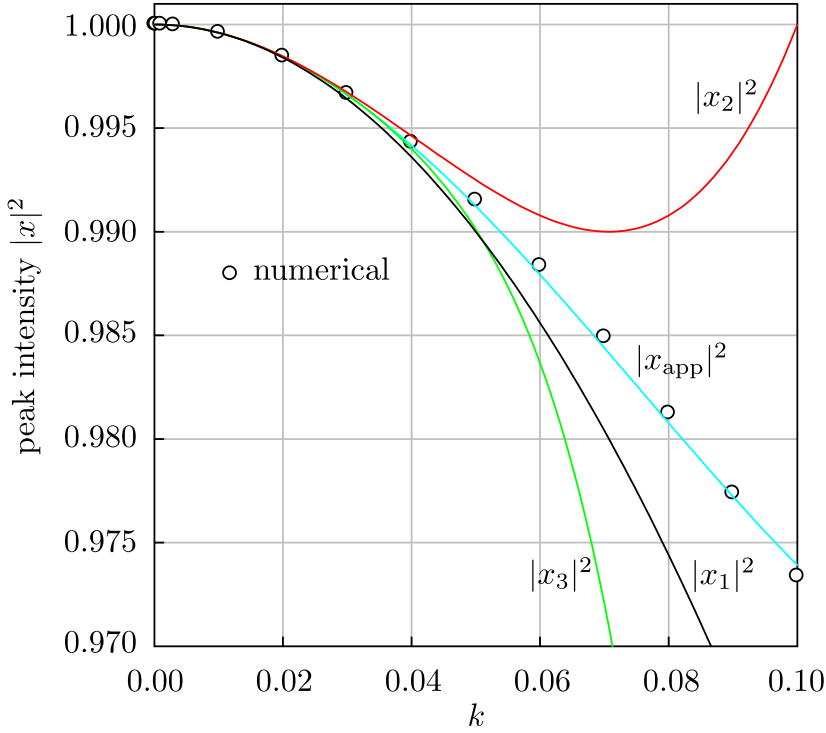}
\caption{\label{fig5}Reduction of peak intensity with increasing $k$ parameter. The black $(|x_1|^2)$,
red $(|x_2|^2)$ and  green $(|x_3|^2)$ lines in the figure show the successive Taylor approximants to
the series, Eq.~\ref{x2pser}: truncated after the first, third, fifth and seventh powers.
 }
\end{figure}

The points in Fig.~\ref{fig5} are the peak intensities  corresponding to different $k$ values, found
directly from numerical solution of Eq.~\ref{dx2=0}. The lines in the figure show the successive approximants
to the series, Eq.~\ref{x2pser}.
However the deviation from quadratic is better-described by a term in $k^{7/2}$; this is shown by the
cyan line in the figure.
Its equation is
\begin{equation*}
\left| {\tilde x_\mathrm{app} \left( {{\tau _{\mathrm{p}}}} \right)} \right|^2 = 1 - 4{k^2} + 44{k^{7/2}}.
\end{equation*}

We note that the peak attenuation is a \emph{second-order} effect; by contrast the frequency shift is
a \emph{first-order} effect.

\section{The parameter $k$}
The fundamental equation of motion, Eq.~\ref{difeq1} involves the quantities $\omega_{\mathrm{z}}$,
$Q$ and $\alpha$. However, as we have seen from Eq.~\ref{erfcsol}, in describing the shape of the resonance
profile these quantities coalesce into the single dimensionless parameter $k$. We note, parenthetically,
that this collapse cannot be demonstrated directly from the original equation of motion; it arises directly
from the high-Q assumption, by which Eq.~\ref{erfcsol} follows from Eq.~\ref{x_hat}.

Within the framework of this approximation the absolute frequency of the oscillator, $\omega_{\mathrm{z}}$
disappears from consideration. We have the width of the resonance $\Delta\omega=\omega_{\mathrm{z}}/Q
$ and the only other quantity in the equation of motion is the sweep rate $\alpha $, which has the dimensions
of frequency squared. The dimensionless $k $ is simply the ratio $\alpha/2(\Delta\omega)^2$.

From the previous section we found, in the slow-sweep limit, that
\begin{equation*}
k=\frac{\delta \omega}{2\Delta\omega};
\end{equation*}
$ k$ is half the shift, expressed as a fraction of the width.

Yet another aspect of the meaning of $k $ will be seen in the next section, where we will find that it
corresponds to the aspect ratio of the system's Wigner distribution.

\section{Wigner representation}\label{section_wigner}
\begin{figure}
\includegraphics[scale=0.9]{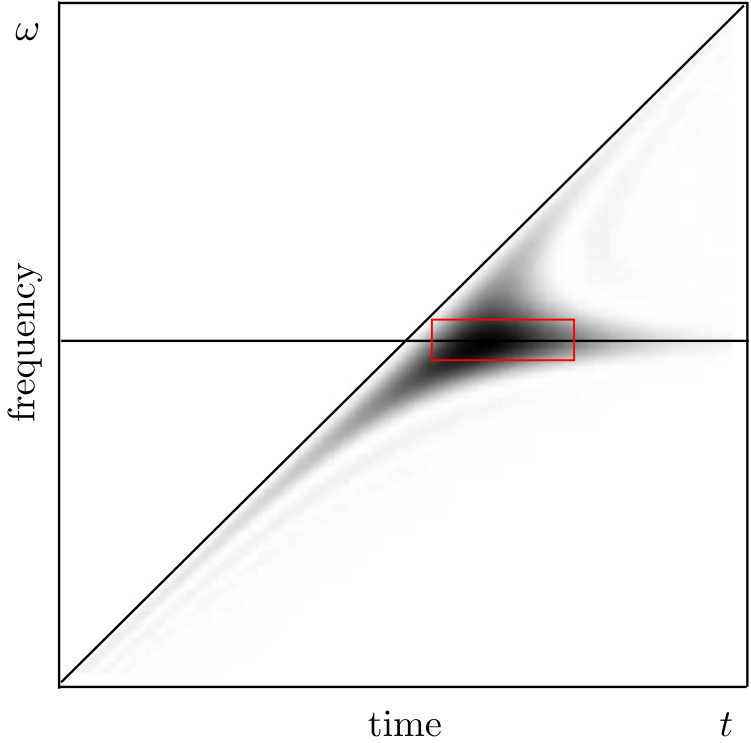}
\caption{\label{fig6}Wigner distribution density for Gliding Tone problem. The horizontal line indicates
the natural resonance of the oscillator $\omega_{\mathrm{z}}$ and the sloping line represents the increase
of drive frequency with time ${\omega } = {\omega _{\mathrm{z}}} + \alpha t$. The significance of the
rectangular box is discussed around Eq.~\ref{-kbox}. }
\end{figure}

The Wigner function attempts to describe how the frequency spectrum of a function -- our $x(t)$ -- varies
with time. The Wigner distribution is defined as~\cite{Wigner1932}
\begin{equation*}
W\left( {t,\omega } \right) = \frac{1}{{2\pi }}\int {{x^*}\left( {t - {\tau  \mathord{\left/
 {\vphantom {\tau  2}} \right.
 \kern-\nulldelimiterspace} 2}} \right){e^{ - {\mathrm{i}}\omega \tau }}x\left( {t + {\tau  \mathord{\left/
 {\vphantom {\tau  2}} \right.
 \kern-\nulldelimiterspace} 2}} \right){\mathrm{d}}\tau }.
\end{equation*}
The interpretation of such a time-frequency distribution is that the integral
\begin{equation}\label{wig_prob}
p = \int\limits_{{t_1}}^{{t_2}} {\int\limits_{{\omega _1}}^{{\omega _2}} {W\left( {t,\omega } \right)}
\,{\mathrm{d}}\omega \,{\mathrm{d}}t} 
\end{equation}
is proportional to the energy in the signal in the frequency range ${\omega _1} < \omega  < {\omega _2}$
 during the time interval ${t_1} < t < {t_2}$. The Uncertainly Principle implies that this result will
be meaningless if the observed cell in time-frequency space is too small; in order for Eq.~\ref{wig_prob}
to be meaningful we require
\begin{equation*}
\left( {{t_2} - {t_1}} \right)\left( {{\omega _2} - {\omega _1}}\right) \gtrsim 1.
\end{equation*}         
It follows from this that there is considerable latitude in the specification of a time-frequency distribution
function; indeed, as Cohen has shown~\cite{Cohen1995}, the Wigner distribution is but one of a large class
of possible time-frequency distributions.
Galleani and Cohen evaluated exactly the Wigner distribution for the Gliding Tone problem~\cite{Galleani2000}.
They found:
\begin{equation*}
W\left( {t,\omega } \right) = \theta \left( \tau  \right)\frac{{{e^{ - {\omega _{\mathrm{z}}}\tau /Q}}}}{{2\alpha
{\omega _{\mathrm{f}}}}}\left\{ {\frac{{\sin 2\left( {\omega  - {\omega _{\mathrm{f}}}} \right)\tau }}{{\omega
\left( {\omega  - {\omega _{\mathrm{f}}}} \right)}} - \frac{{\sin 2\left( {\omega  + {\omega _{\mathrm{f}}}}
\right)\tau }}{{\omega \left( {\omega  + {\omega _{\mathrm{f}}}} \right)}}} \right\}
\end{equation*}
where
\begin{equation*}\tau  = t - {\omega  \mathord{\left/
 {\vphantom {\omega  \alpha }} \right.
 \kern-\nulldelimiterspace} \alpha }\end{equation*} 
and $\omega_{\mathrm{f}}$ is defined in Eq.~\ref{wf}. 
A typical density plot is shown in Fig.~\ref{fig6}.
The horizontal line indicates the natural resonance of the oscillator $\omega_{\mathrm{z}}$ and the sloping
line represents the increase of drive frequency with time $\omega = {\omega _{\mathrm{z}}} + \alpha t$.
The Wigner distribution function indicates that the response is concentrated close to the intersection
of these lines -- where the excitation is in the vicinity of the natural resonance. The density along
the sloping line shows the response at the excitation frequency. The density along the horizontal line
shows the ``ringing'' at the oscillator's natural frequency caused by the ``shock'' of the excitation frequency
varying. The  response being all to the right of the sloping line is a manifestation of causality. 

The height of the distribution, the extension in frequency, is the breadth of the resonance, $\omega_{\mathrm{z}}/Q$.
The width of the distribution, the extension in time, is the ring-down time ${\tau _{\mathrm{d}}} = 2{Q
\mathord{\left/
 {\vphantom {Q {{\omega _{\mathrm{z}}}}}} \right.
 \kern-\nulldelimiterspace} {{\omega _{\mathrm{z}}}}}$. This translates to a frequency \emph{sweep} width
of $2{{\alpha Q} \mathord{\left/
 {\vphantom {{\alpha Q} {{\omega _{\mathrm{z}}}}}} \right.
 \kern-\nulldelimiterspace} {{\omega _{\mathrm{z}}}}}$. And the ratio of these frequencies is  $2\alpha
{\left( {{Q \mathord{\left/
 {\vphantom {Q {{\omega _{\mathrm{z}}}}}} \right.
 \kern-\nulldelimiterspace} {{\omega _{\mathrm{z}}}}}} \right)^2}$; this is precisely Barber and Ursell's
$k$  parameter~\cite{Barber1948}, Eq.~\ref{kdef}. Thus $k$ determines the \emph{aspect ratio} of the
Wigner distribution:
\begin{equation}\label{-kbox}
k = \frac{{{\mathrm{extension~in~time~space}}}}{{{\mathrm{extension~in~frequency~space}}}},
\end{equation}
with both measured in the same units. A box of the appropriate aspect ratio is shown in Fig.~\ref{fig6}.
Observe that it does suggest the aspect ratio of the Wigner density. 

The above result also indicates the possibility that the Wigner distribution may be expressible in universal
form. In order to demonstrate this universality we take the Wigner distribution as
\begin{equation*}
W\left( {t,\omega } \right) = \frac{1}{{2\alpha {\omega _{\mathrm{f}}}}}\theta \left( \tau  \right){e^{
- {{{\omega _{\mathrm{z}}}\tau } \mathord{\left/
 {\vphantom {{{\omega _{\mathrm{z}}}\tau } Q}} \right.
 \kern-\nulldelimiterspace} Q}}}\frac{{\sin 2\left( {\omega  - {\omega _{\mathrm{f}}}} \right)\tau }}{{\omega
\left( {\omega  - {\omega _{\mathrm{f}}}} \right)}}
\end{equation*}
where we have used the high-Q procedure of discarding the negative frequency resonance. We first apply
a horizontal shear to this function so that $\tau=t-\omega/\alpha$ becomes the independent ``time'' variable;
the sloping line of Fig.~\ref{fig6} becomes the vertical axis. And then we scale the time and frequency
variables to the dimensionless\begin{equation*}
\Omega  = \omega \sqrt {{k \mathord{\left/
 {\vphantom {k \alpha }} \right.
 \kern-\nulldelimiterspace} \alpha }} ,\;\;\;\;\;\;T = \tau \sqrt {{\alpha  \mathord{\left/
 {\vphantom {\alpha  k}} \right.
 \kern-\nulldelimiterspace} k}}.
\end{equation*}
We also need the auxiliary ${\Omega _{\mathrm{f}}} = {\omega _{\mathrm{f}}}\sqrt {{k \mathord{\left/
 {\vphantom {k \alpha }} \right.
 \kern-\nulldelimiterspace} \alpha }} $. In terms of these variables the Wigner distribution function
is\begin{equation*}
W\left( {T,\Omega } \right) = \frac{{\sqrt k }}{{2{\alpha ^{{3 \mathord{\left/
 {\vphantom {3 2}} \right.
 \kern-\nulldelimiterspace} 2}}}{\omega _{\mathrm{f}}}\omega }}\theta \left( T \right){e^{ - \sqrt 2
T}}\frac{{\sin \left( {\Omega  - {\Omega _{\mathrm{f}}}} \right)T}}{{\left( {\Omega  - {\Omega _{\mathrm{f}}}}
\right)}}.\end{equation*}
In the spirit of the high-Q approximation we argue that $W\left( {T,\Omega } \right)$ is non-zero only
in the vicinity of  $\omega \sim{\omega _{\mathrm{f}}}$, and we thus write\begin{equation*}
W\left( {T,\Omega } \right) = \frac{{\sqrt k }}{{2{\alpha ^{{3 \mathord{\left/
 {\vphantom {3 2}} \right.
 \kern-\nulldelimiterspace} 2}}}\omega _{\mathrm{f}}^2}}\theta \left( T \right){e^{ - \sqrt 2 T}}\frac{{\sin
\left( {\Omega  - {\Omega _{\mathrm{f}}}} \right)T}}{{\left( {\Omega  - {\Omega _{\mathrm{f}}}} \right)}}.
\end{equation*}   
There is a system-dependent pre-factor but the time and frequency dependence are all in the dimensionless
variables. 
\begin{figure}
\includegraphics[scale=1.0]{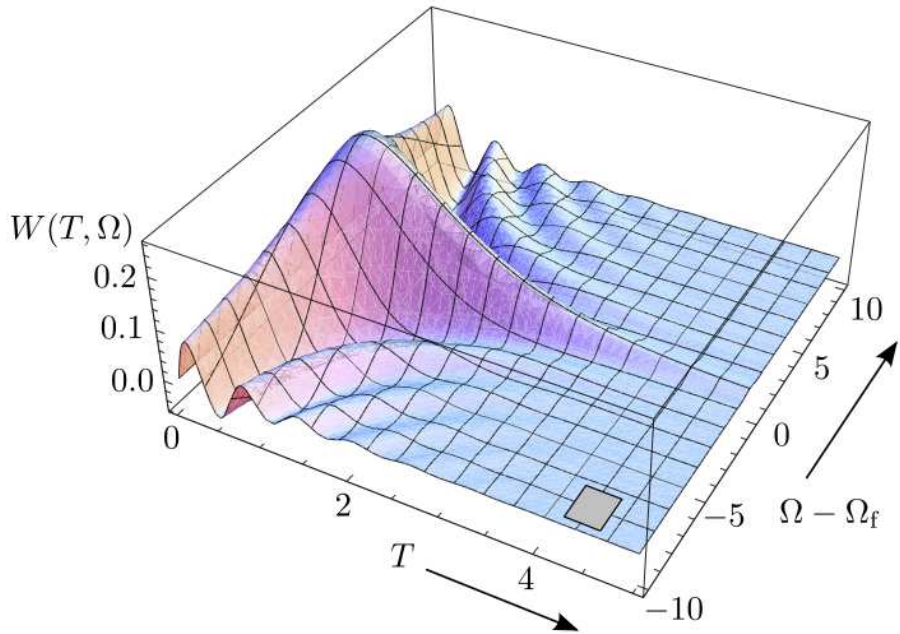}
\caption{\label{fig7}Wigner distribution in terms of reduced variables}
\end{figure}
This is plotted in Fig.~\ref{fig7}. The oscillations (and in particular the fact that $W\left( T,\Omega
\right)$ goes negative) are an artefact of the Wigner distribution; it is only the integral over an ``uncertainty''
area that is meaningful as a (necessarily non-negative) probability. We show, in the figure, the grey
rectangle indicating the scale of this region.

\section{Complex plots}
A convenient graphical representation of a system's response is provided by plotting the imaginary part
against the real part of the complex response function, sometimes referred to as a \emph{hodograph}.
In the context of an oscillator the parametric plot is mapped out as as one sweeps through the resonance,
the applied frequency being the variable parameter of the plot. In comparison with separate plots of
the real and imaginary parts of the response \emph{versus} excitation frequency, the complex plot involves
a loss of information: frequency is no longer an explicit variable. However this can an advantage.  In
general the response function of a system will be complicated. In its mathematical description the dependence
on frequency typically will be as a product $\omega \tau_{\mathrm{c}}$, where $\tau_{\mathrm{c}} $ is
a characteristic time. It is then equally valid to regard $\omega \tau_{\mathrm{c}}$ as the parameter
of the plot. The actual frequency and characteristic time are not important; it is only their product.
Thus the parametric plot is able to demonstrate \emph{generic} properties of the response function. Indeed
one might also measure the response at constant applied frequency while varying $\tau_{\mathrm{c}}$,
perhaps indirectly by changing the temperature or pressure.   

The complex plot of the quasi-static or normal Lorentzian resonance has a circular locus. Distortion
of the resonance, as follows from the Gliding Tone effect, manifests itself as a deviation from circularity.
Some examples are shown in Fig.~9 (where the curves are traversed in the anti-clockwise sense). The distortion
worsens with increasing $k$ parameter, as one would expect.  There are three effects to note: i) before
resonance the magnitude response (the radius vector) is smaller than the quasistatic value; ii) on passing
the resonance the response will become greater, leading, for large $k$ to iii) ringing.    
\begin{figure}
\includegraphics[scale=1]{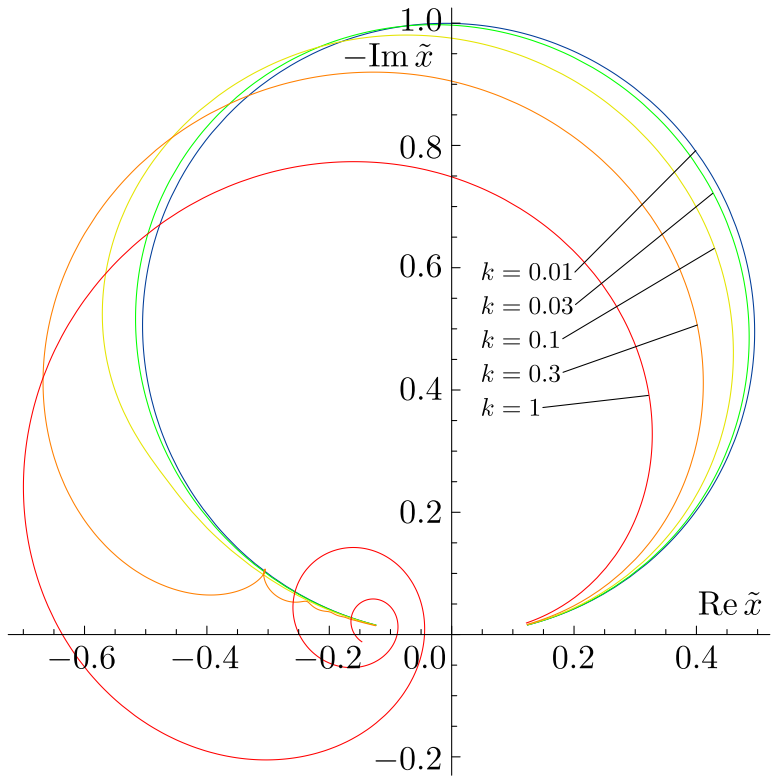}
\caption{\label{fig7}Complex plot of oscillator response for different $k$ parameters}
\end{figure}

\section{Experimental measurements}
The motivation for the above study has been the interpretation and understanding of our experiments on
the putative supersolid phase of helium. Torsional oscillators of high Q-factor provide sensitive sensors
of ``lost mass'' through its reflection in small changes of resonance frequency. Indeed such an oscillator
was used in the discovery that the superfluid transition in two-dimensional liquid $^4$He was of the
Kosterlitz-Thouless type~\cite{Bishop1978}. 

We now show some measurements performed in our laboratory, on torsional oscillators as typically used
in low-temperature physics research~\cite{Richardson1998}. The oscillator -- particularly the torsion
rod -- is made of coin silver, and carefully annealed to increase its Q-factor. The oscillator is driven
and the response detected both electrostatically. Our measurements were made at low temperatures, where
$Q$s of order $10^6$ are achievable.  

\begin{figure}
\includegraphics[scale=0.7]{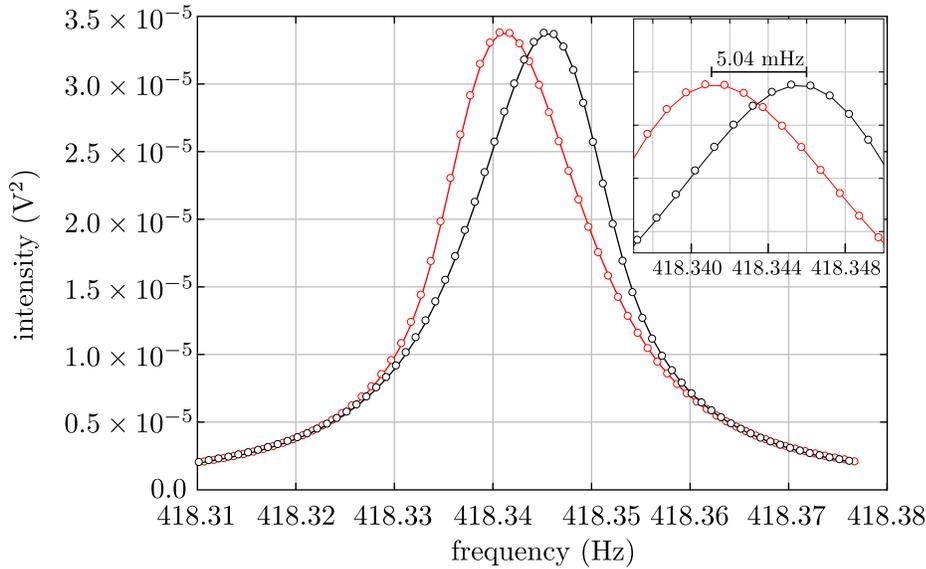}
\caption{\label{GlideLow}Distortion of resonance caused by a finite sweep rate corresponding to a $k$
parameter of 0.081. Black: up sweep, Red: down sweep.
 }
\end{figure}

A first example is given in Fig.~\ref{GlideLow}. The plots show sweeps up and down through the resonance
and the displacement of the peaks indicates that there is an effect which must be addressed in order
to draw correct inferences from frequency measurements. The frequency was swept at $7.3\times10^{-5}
\mathrm{ Hz \hspace{2pt}s^{-1}}$ up and down through the resonance at $418.343510\,\mathrm{Hz}$, the
oscillator having a Q-factor of 24,750. According to Eq.~\ref{kdef} this gives a $k$ parameter of 0.081,
implying a peak's (dimensionless) time shift $\tau = 0.149$ which from Eq.~\ref{deltaomega}, corresponding
to a frequency shift of 2.52~mHz. The bar in the inset shows the expected separation of 5.04~mHz between the
experimental peaks.

\begin{figure}
\includegraphics[scale=1]{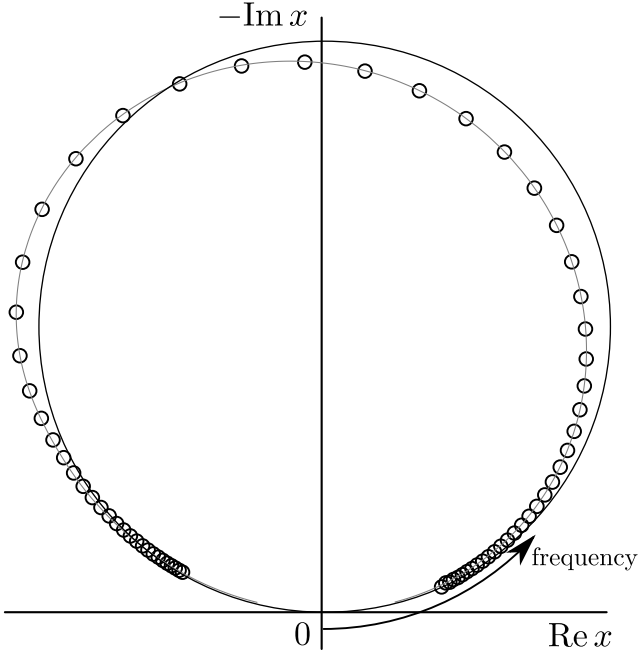}
\caption{\label{hodo2}Distortion of complex plot caused by a finite sweep rate corresponding to a $k$
parameter of 0.081. For comparison the corresponding quasi-static circular profile is shown.   }
\end{figure}

The hodograph corresponding to this plot (the up sweep only) is shown in Fig.~\ref{hodo2}. It is salutary
to note that while either single sweep of Fig.~\ref{GlideLow} might not be recognized as embodying a
measure of distortion -- indeed a Lorentzian fit through the data may be misleadingly good -- it is the
complex plot that gives the clearest indication of resonance distortion.

\begin{figure}
\includegraphics[scale=0.75]{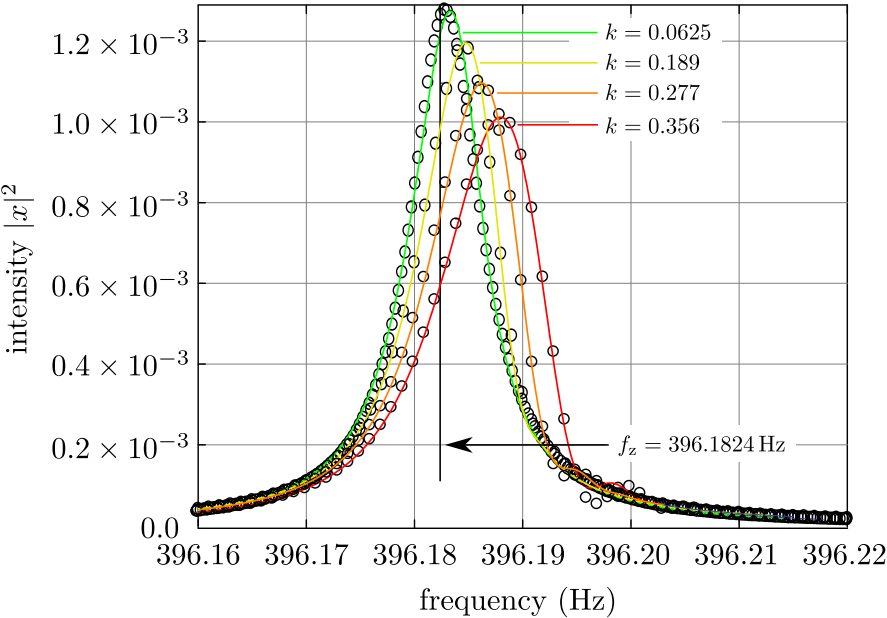}
\caption{\label{k-lot}Resonance intensity for a range of $k$ values  }
\end{figure}

Finally, in Fig.~\ref{k-lot} we show a sequence of spectra, taken with different sweep rates. The quasi-static
resonance frequency is 396.1824~Hz. This is shown in the figure and already the slowest curve ($k=0.0625$)
shows a discernible shift. As $k$ increases we see evolution of the characteristic features: i) shift
of the peak frequency, ii) reduction in the peak height, iii) increasing asymmetry of the curve, and
iv) the emergence of ringing at the fastest sweep. The lines in the figure are calculated from Eq.~\ref{erfcsol}.
The agreement is gratifying.

\section{Conclusion}
The solution to the Gliding Tone problem has been obtained in closed form in terms of known functions.
The sole non-exact feature is that the Q-factor of the oscillator is assumed to be high. In the high-Q
case the variables of the problem combine into a single parameter ($\gamma$), in terms of which the solution
is expressed. We investigated the deviation from the quasi-static oscillator response as a consequence
of the finite sweep rate, showing that the resonance distortion depended on the dimensionless parameter
$k$. To lowest order the delay in the appearance of the resonance peak corresponds to double the oscillator's
ring-down time. Experimental observations on a high-Q torsional oscillator are shown to be consistent
with the theory presented. 

\section{Acknowledgements}
We thank Dr. Jan Ny\`eki for raising this problem and Prof.~Leon Cohen for expressing encouraging interest
in the work and for providing some helpful contextual comments. Prof. Lorenzo Galleani was kind enough
to clarify the form of the Gliding Tone Wigner distribution. We are grateful to Prof.~Nico Temmi for
advice on the nomenclature and properties of the ``error-related'' functions. 

We used the \textit{Mathematica} software package for performing many of the algebraic manipulations;
we note, in particular, that the mathematical steps in arriving at Eq.~\ref{sertau} were exceedingly
labour-intensive as sufficient terms in the intermediate series must be taken in order for the contributions
to each power of $k$ to be exhausted. It took 8.5 hours on a fairly standard PC to obtain terms up to
$k^7$ and 128 hours for the next term!

\end{document}